\documentclass[prb,twocolumn]{revtex4}
\usepackage{graphicx}
\usepackage{times}
\usepackage{setspace}
\usepackage{wrapfig}
\usepackage{epstopdf}

\newcommand{\units}[1]{\ensuremath{\mathrm{#1}}}
\newcommand{\amount}[2]{\ensuremath{#1\:\units{#2}}}

\begin{document}
\title{Charge sensing and controllable tunnel coupling in a Si/SiGe double quantum dot}
\author{C. B. Simmons}
\author{Madhu Thalakulam}
\author{B. M. Rosemeyer}
\author{B. J. Van Bael}
\author{E. K. Sackmann}
\author{D. E. Savage}
\author{M. G. Lagally}
\author{R. Joynt}
\author{M. Friesen}
\author{S. N. Coppersmith}
\author{M. A. Eriksson}
\affiliation{University of Wisconsin-Madison, Madison, Wisconsin 53706, USA}
\begin{abstract}
We report integrated charge sensing measurements on a Si/SiGe double quantum dot.  The quantum dot is shown to be tunable from a single, large dot to a well-isolated double dot.  Charge sensing measurements enable the extraction of the tunnel coupling $t$ between the quantum dots as a function of the voltage on the top gates defining the device.  Control of the voltage on a single such gate tunes the barrier separating the two dots.  The measured tunnel coupling is an exponential function of the gate voltage.  The ability to control $t$ is an important step towards controlling spin qubits in silicon quantum dots.
\end{abstract}
\maketitle

Following a proposal to use gated quantum dots to host spin qubits in semiconductors,\cite{Loss:1998p1165} a series of recent experiments using GaAs/AlGaAs heterostructures have demonstrated the measurement and manipulation of quantum dot semiconductor spin qubits.\cite{Elzerman:2004p729,Petta:2005p184,Johnson:2005p484,Koppens:2006p732,Koppens:2008p1380,PioroLadriere:2008p1634,Reilly:2008p1445}  At low temperatures, the $T_2$ dephasing time for semiconductor spin qubits is usually limited by hyperfine coupling between the electron and nuclear spins.\cite{Khaetskii:2003p1655,deSousa:2003p1660,Yao:2006p1662}  Because Si and Ge have abundant isotopes with zero nuclear spin, a number of proposals for spin qubits in semiconductors have focused on these materials,\cite{Kane:1998p1224,Vrijen:2000p1643,Friesen:2003p65,Morello:2009preprint} and gated quantum dots have now been fabricated in Si/SiGe,\cite{Slinker:2005p49,Sakr:2005p174,Berer:2006p532,Klein:2007p32,Simmons:2007p4} Si-MOS systems,\cite{Simmel:1999p1550,Rokhinson:2001p1655,Fujiwara:2006p1652,Jones:2006p1651,Angus:2007p845,Zimmerman:2007p1527,Liu:2008p751,Lansbergen:2008p1545,Nordberg:2009pc} Si nanowires,\cite{Zhong:2005p1143,Hofheinz:2006p1654,Hu:2007p465,Zwanenburg:2009p1579} and patterned donor layers.\cite{Fuhrer:2009p1549}

One of the important features of quantum dots is that they can be easily coupled to one another in the form of a double quantum dot.  The tunnel coupling $t$ between the two dots in a double dot is an important quantity, because several different qubit operations can be performed by controlling this tunnel rate as a function of time.  For approaches in which the qubit is a single electron spin, control of $t$ enables the performance of SWAP operations between two qubits.\cite{Loss:1998p1165}  In a two-electron triplet-singlet qubit, controlling $t$ enables single-qubit rotations.\cite{Levy:2002p1446}  And in an architecture in which each logical qubit is composed of three spins in three different quantum dots, control of $t$ alone enables universal quantum computation.\cite{Divincenzo:2000p1642}

Here we report measurements of a Si/SiGe double quantum dot with an integrated quantum point contact (QPC) used for charge sensing.  The lateral confinement of the quantum dot is tuned elecrostatically using metal top gates, and the device characteristics allow comprehensive control of the inter-dot tunnel coupling.  We demonstrate a smooth transition from a well-defined double dot with weak tunnel coupling, through a strongly coupled quantum dot molecule, to a single, large quantum dot.  Charge sensing with the QPC is used to probe the transfer of charge from one dot to the other.  The sharpness of this charge transition depends critically on the inter-dot tunnel coupling, and fits to this transition allow the tunnel coupling to be extracted over the range from a weakly to a strongly coupled double dot.  We measure a change in the tunnel coupling of more than a factor of $10$, corresponding to more than a $100$-fold modulation in the exchange coupling between the quantum dots.

\begin{figure}
\includegraphics{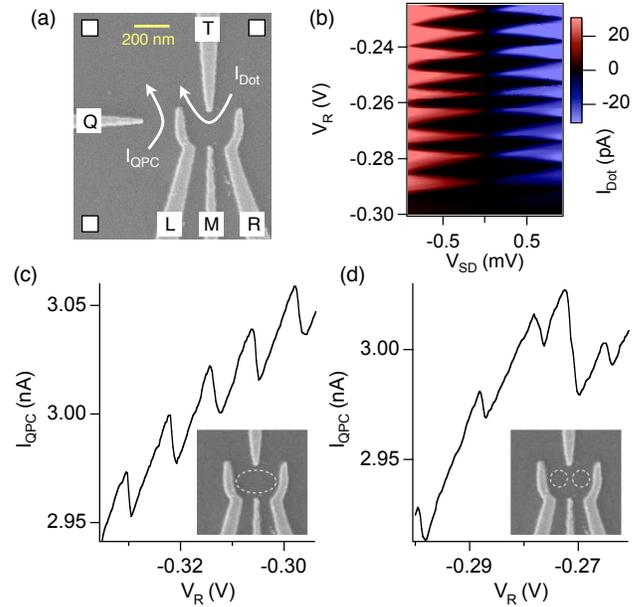}
\caption{(a) Scanning electron micrograph showing the top gate geometry for the measured device.  The transport current through the dot is indicated by $I_{\text{dot}}$ and the charge sensor current by $I_{\text{QPC}}$.  (b) Coulomb diamonds measured in transport through the quantum dot tuned to be a single dot.  (c) Charge sensing signal from the coupled quantum point contact showing the periodic steps in $I_{\text{QPC}}$ resulting from charge transitions on the neighboring single quantum dot, shown schematically by the dashed oval superimposed on the SEM image.  (d) Charge sensing signal from a weakly coupled double dot, showing two different charge sensing amplitudes.}
\end{figure}

A scanning electron micrograph of the device design is shown in Fig.~1a.  The two-dimensional electron system (2DES) is located approximately \amount{75}{nm} below the surface of the heterostructure.  The carrier density and mobility of the 2DES determined from Hall and Shubnikov-de Haas measurements are \amount{5\times10^{11}}{cm^{-2}} and \amount{44,000}{cm^{2}/Vs} respectively.  The four gates labeled $T$, $L$, $M$, and $R$ are used to create the quantum dot confinement, while gate $Q$ in combination with gate $L$ form the electrostatically coupled QPC.  This device design incorporates several features that appear to be important for the results reported here.  First, in order to minimize the dot size, and similar to Ref.~\onlinecite{PioroLadriere:2008p1634}, this device does not have the ``plunger'' gates that are often found in double quantum dot designs.  The result is that each side of the quantum dot is smaller than previously published Si/SiGe quantum dots that have implemented a similar gate geometry.\cite{Sakr:2005p174,Klein:2007p32,Simmons:2007p4}  Second, the side-gates $L$ and $R$ are quite narrow ($\sim40\%$ the width of the side-gate in Ref.~\onlinecite{Simmons:2007p4}) in order to reduce the screening between the dot and the QPC to maximize the sensor response to charge transitions.  As we show below, this quantum dot can be tuned to be a single or a double quantum dot by adjusting the voltage on the middle gate $V_M$.

In the single dot regime, with small $|V_M|$, symmetric Coulomb diamonds are observed in the transport current through the dot $I_{\text{dot}}$ as a function of the source-drain voltage $V_{\text{SD}}$ and $V_R$, as shown in Fig.~1b.  The last measurable diamond gives a single-dot charging energy $E_c=e^2/C = 880 \mu eV$ and a capacitance between gate $R$ and the dot of \amount{17}{aF}.  The charging energy corresponds to a total dot capacitance \amount{C = 182}{aF}, and this gives a lever arm $C_R/C = 0.092$.  After this last diamond, the tunnel barriers between the dot and the leads are too opaque to identify further charge transitions using transport current through the dot, and charge sensing must be used.

Figure~1c and d show the use of a nearby QPC for non-invasive measurement of the charge states of the quantum dot(s).  As $V_R$ is made more negative, electrons are removed from the dot, and the resulting decrease in the electrostatic potential of the dot causes a sudden increase in $I_{\text{QPC}}$.  In addition, there is an overall linear decrease in $I_{\text{QPC}}$ due to the direct coupling between gate $R$ and the QPC, resulting in the saw tooth pattern of $I_{\text{QPC}}$ vs.\ $V_R$ shown in Fig. 1c.  The uniform height and periodic steps in $I_{\text{QPC}}$ indicate the presence of a single quantum dot.  When the dot is tuned into a weakly coupled double dot, the charge sensor reports two distinct amplitudes for transitions on the left and right dots, as seen in Fig.~1d.  The left dot is physically closer to the QPC, and it produces a larger step \amount{\delta I \sim 50}{pA}, while a one electron change on the right dot yields \amount{\delta I \sim 15}{pA}.  These charge signals on the QPC correspond to $\sim 7\times10^{-3}~e^2/h$ and $2\times10^{-3}~e^2/h$ for the left and right dot respectively, in good agreement with observations in similar systems.\cite{Dicarlo:2004p1440}

\begin{figure}
\includegraphics{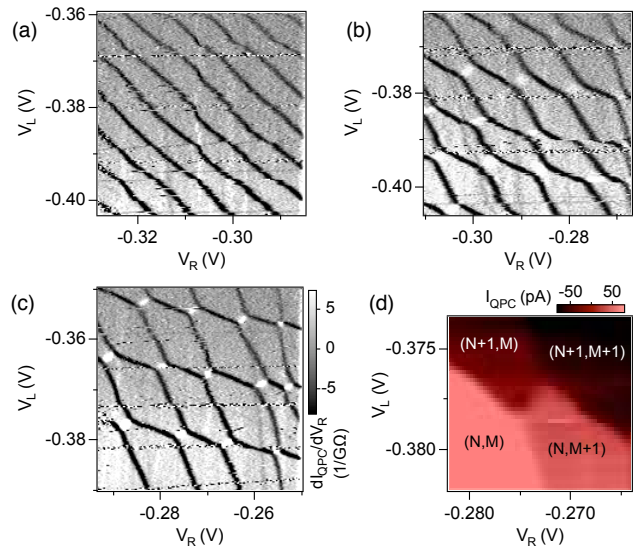}
\caption{(a)-(c) $dI_{\text{QPC}}/dV_R$ measured as a function of $V_R$ and $V_L$ for increasing $|V_M| $ (-0.5 V, -0.7 V and -1.0 V) (gray-scale shown in (c)).  Three regimes of the quantum dot are evident from the dark transition lines: a single, large dot, a strongly coupled quantum dot molecule, and a weakly coupled double dot.  (d) $I_{\text{QPC}}$ at a single vertex of the honeycomb diagram with a background plane removed, showing four charge states of the double dot.}
\end{figure}

\begin{figure}
\includegraphics{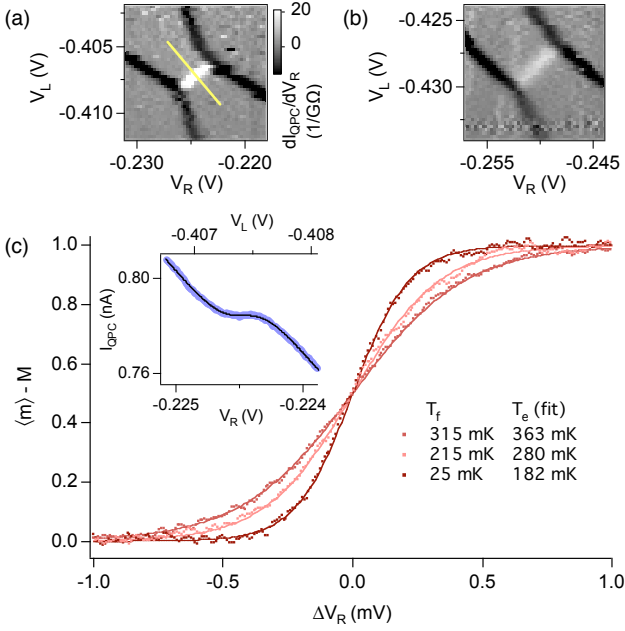}
\caption{$dI_{\text{QPC}}/dV_R$ measured as a function of $V_R$ and $V_L$ showing a single vertex of the charge stability diagram for the double quantum dot in the weak (a) and strong (b) coupling regimes (gray-scale shown in (a)).  The superimposed yellow line in (a) indicates the detuning diagonal for the $(N,M+1)$ and $(N+1,M)$ charge states. (c) Inset: $I_{\text{QPC}}$ along the detuning diagonal of a weakly coupled double dot with the corresponding fit to the interdot transition given by Eq.~(\ref{IQPC}).  Main plot: The interdot transition rescaled into units of excess charge on the right dot at three different temperatures.  The solid curves are fits using Eq.~(\ref{m}) with $t=0$.  Each of the data sets corresponds to an average of 15 sweeps.}
\end{figure}

Figure~2a-c show the transition from a single dot to a well-separated double dot, as indicated by the charge sensing signal $dI_{\text{QPC}}/dV_R$ as a function of $V_R$ and $V_L$, taken at three increasingly negative values of the voltage on the middle gate $V_M$.  The numerical differentiation of $I_{\text{QPC}}$ as a function of $V_R$ highlights the charge transitions on the dot which appear as black or white lines in these plots.  In Fig.~2a, with \amount{V_M = -0.5}{V}, the charge transitions form a single set of parallel lines, indicating that the system is composed of a single, large quantum dot.  The charge transitions are nearly symmetric with respect to $V_R$ and $V_L$, implying that the single quantum dot is coupled roughly equally to gate $R$ and gate $L$, as expected based on the symmetric lithographic gate design (Fig.~1a).  By making the voltage on gate $M$ more negative (\amount{V_M=-0.7}{V}), the single dot is split into a strongly coupled double dot, where the left side is coupled dominantly to gate $L$ and the right side to gate $R$, as shown by the bending of the transition lines in Fig.~2b.  Further decreasing $V_M$ to \amount{-1.0}{V} results in the weak-coupling regime and a well-isolated pair of quantum dots (Fig.~2c).

The charge transition lines in Fig.~2c form the expected honeycomb pattern where each cell of the honeycomb corresponds to well-defined electron occupations of the left and right dots.\cite{VanDerWiel:2002p1382}  Previous observation of double dot behavior in Si/SiGe was made in a structure that was designed to be a single quantum dot, and numerical modeling was required to understand how double dot behavior could arise in such a structure.\cite{Shaji:2008p1372}  In contrast, the structure used here and the data shown in Fig.~2 lead to a natural identifcation of the dot locations: from the slopes and the spacings of the charge transition lines in Fig.~2c, we find that the left dot is coupled $\sim2.2$ times more strongly to gate $L$ than gate $R$ (the more horizontal transitions), while the right dot is coupled $\sim2.0$ times more strongly to gate $R$ than gate $L$ (the more vertical transitions).

Figure~2d shows $I_{\text{QPC}}$ at one of the vertices of the honeycomb stability diagram with a background plane removed.  Four distinct values of $I_{\text{QPC}}$ are visible, corresponding to four different charge states on the double quantum dot.  Moving an electron from the left dot to the right dot results in an increase in $I_{\text{QPC}}$.  As shown in Fig.~3a and b, this increase appears as a white line in a plot of $dI_{\text{QPC}}/dV_{R}$.  In Fig.~3a, with \amount{V_M=-1.0}{V}, the transition appears as a sharp white line, indicating a sharp interdot transition and weak tunnel coupling between the dots.   In Fig.~3b, \amount{V_M=-0.825}{V} and the transition appears as a broad and less intense white line, indicating strong tunnel coupling and increased charge delocalization between the dots.

Moving along the detuning diagonal in gate voltage space, indicated by the superimposed yellow line in Fig.~3a, causes a single electron to shift from one dot to the other.  Measurements of this transition allow the extraction of a number of important quantities describing the charge sensor.  The inset of Fig.~3c shows $I_{\text{QPC}}$ measured as $V_L$ and $V_R$ are simultaneously swept to follow the detuning diagonal.  On top of a linear background, due to the direct gate-QPC coupling, there is a step in $I_{\text{QPC}}$ when \amount{V_R\approx-0.2245}{V} due to the transition between the $(N,M+1)$ and the $(N+1,M)$ charge states.

Following the method of DiCarlo et al.,\cite{Dicarlo:2004p1440} we treat the $(N,M+1)$ and $(N+1,M)$ charge states as a two level system and fit the effect of the ground charge state transition on $I_{\text{QPC}}$ using:
\begin{equation}\label{IQPC}
I_{QPC}(\epsilon) = I_0 + \delta I \frac{\epsilon}{\Omega}\tanh \left( \frac{\Omega}{2k_{B}T_{e}} \right)
+ \frac{\partial I}{\partial \epsilon}\epsilon
\end{equation}
Here, $\epsilon$ is the detuning energy, $T_e$ is the electron temperature, and $\Omega$ is the ground and excited state energy splitting, $\Omega = \sqrt{\epsilon^2+4t^2}$, where $t$ is the tunnel coupling.  $I_0$, $\delta I$, and $\partial I/\partial \epsilon$ are the background current, the sensor sensitivity, and the direct gate-QPC coupling, respectively.  These three parameters describe the system and are important for the extraction of the tunnel coupling below.  The detuning energy $\epsilon$ is related to the gate voltage $V_R$ by a lever arm $\alpha$,  $\epsilon=\alpha V_R$.  As shown by the solid line in the inset of Fig.~3c, fits using Eq.~\ref{IQPC} provides good agreement with the data.  Using the results for the three fit parameters that characterize the QPC sensor, $I_{\text{QPC}}$ can be converted into units of excess charge on the right dot, $\langle m\rangle-M$, and we report the data in the main panel of Fig.~3a and in Fig.~4 below in this way.

\begin{figure}
\includegraphics{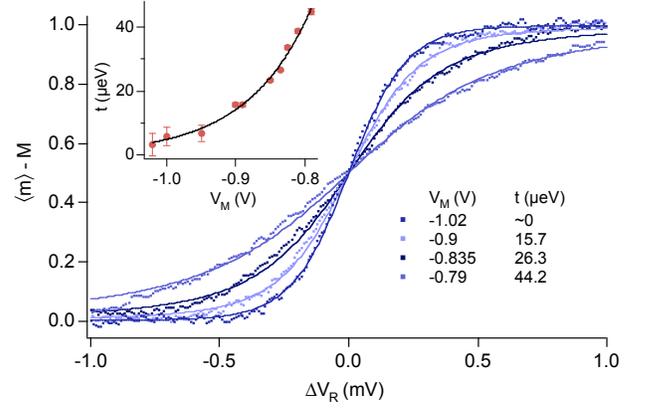}
\caption{Excess charge on the right dot along the interdot transition taken with four different values of the double dot tunnel coupling $t$.  Each data set is an average of 15 sweeps, and the solid lines are fits to the data using Eq.~(\ref{m}).  Inset: The extracted values of $t$ as a function of $V_M$.  Solid line is an exponential fit to the data.}
\end{figure}

When plotted versus gate voltage, the width of the interdot charge transition is dependent on three factors: $T_e$, $\alpha$, and $t$.  To determine $T_e$ and $\alpha$, we tune the double dot into the weak coupling regime, where $t\approx 0$, by making $V_M$ increasingly negative until the transition is broadened only by temperature.  In this limit, and with $V_M$ fixed, raising $T_e$ by raising the temperature of the dilution refrigerator broadens out the transition, as shown in Fig.~3c.  At low temperatures, the electron temperature is warmer than the refrigerator temperature, due to the balance between the cooling power supplied to the electron system and heating from external sources.  At higher temperatures, the electron temperature $T_e$ approaches the refrigerator temperature $T_f$.   We fit the temperature dependence with a simple relationship between $T_e$ and $T_f$, $T_e = \sqrt{T^2_f + T^2_0}$, where $T_0$ arises due to the effect of external heating.  The parameters $T_0$ and $\alpha$ are determined by simultaneously fitting curves of the interdot transition in the weak coupling limit ($t=0$ and $\epsilon = \Omega$) at six different $T_f$ using the formula,
\begin{equation}\label{m}
\langle m\rangle-M = \frac{1}{2}\left( 1 + \frac{\epsilon}{\Omega}\tanh \left( \frac{\Omega}{2k_B\sqrt{T^2_f + T^2_0}} \right) \right).
\end{equation}
Three of the datasets and the corresponding fits are shown in Fig.~3c.  We find $\alpha = 0.145$ and \amount{T_0 = 180}{mK}, corresponding to \amount{T_e=182}{mK} at the base temperature of the dilution refrigerator.

To determine $t$ as a function of $V_M$, the extracted $T_e$ and $\alpha$ are used to fit curves of the interdot charge transition.  Fig.~4 shows four such fits to the experimental data, with $t$ as the sole free parameter.  At small $|V_M|$, the increased tunnel coupling results in a clear broadening of the interdot charge transition, corresponding to increasingly large $t$.

The largest uncertainty in the values for $t$ extracted by this method arise from the uncertainty in the determination of $\alpha$ and $T_e$.  As described above, $\alpha$ and $T_e$ were determined from fits to the thermally broadened curves based on the assumption $t=0$ in the weak coupling limit.  However, a good fit to those curves can be achieved with $t$ as high as \amount{6}{\mu eV}, resulting in an uncertainty of 3\% in $\alpha$ and 12\% in $T_e$.  The corresponding effect on the uncertainty in $t$ is significant for the smallest $t$, while for large $t$ the uncertainty is very small.

We plot the resulting $t$ as a function of $V_M$ in the inset of Fig.~4.  The data are well fit by an exponential curve.  Changing $V_M$ by \amount{\sim200}{mV} changes $t$ by more than a factor of $10$, corresponding to tuning the exchange interaction $J$ by more than a factor of $100$.\cite{Loss:1998p1165}  In analogy with predictions of oscillations in the exchange coupling between donors,\cite{PhysRevLett.88.027903} it is conceivable that $t$ (and thus $J$) could oscillate as a function of $V_M$, due to the multiple valleys in Si.  As shown in the inset to Fig.~4,  within the experimental uncertainty we observe no indication of such an oscillation in the tunnel coupling $t$ as a function of $V_M$.  This smooth behavior could indicate either an effectively smooth quantum well interface in the active region of the device,\cite{Friesen:2006p45} as consistent with measurements of a large valley splitting in a QPC,\cite{Goswami:2007p35} or it could be caused by valley relaxation in the presence of a rough interface.\cite{Friesen:2009unpublished}

Another important characteristic of quantum dots is their charge stability and noise behavior.  In Fig. 2a-c, horizontal stripes of noise cut across the charging diagrams at several points.  Based on their behavior as a function of the voltages on the gates defining the quantum dots and the QPC, we believe this noise is due to one or more charge traps beneath gate $Q$.   As is clear from Fig.~2, changing the voltage on gate $L$ tunes the noise source in and out of resonance, and this behavior is consistent with previous reports that attributed some types of charge noise to the donor layer in semiconductor heterostructures.\cite{Buizert:2008p1511}  Interestingly, this noise source seems to have little effect on the dots themselves.  As can be seen in the figures, the lines corresponding to transitions in the quantum dot charge states pass through these noise stripes for the most part unchanged.  The quantum dots themselves showed excellent long term stability, with charge drift less then 0.03$e$ (which was the scan resolution) over a 7 day measurement period.  This charge stability is consistent with other Si-based SETs which have shown drift of less than 0.01$e$ over a time period of weeks.\cite{Zimmerman:2007p1527}

In conclusion, we have demonstrated charge sensing measurements on a Si/SiGe quantum dot using an integrated quantum point contact.  We have shown that the dot can be tuned into a single or double quantum dot.  The device exhibited good charge stability, and the charge sensor measurements could resolve the interdot transition of a single electron at fixed total charge.  Broadening of the interdot transition caused by charge delocalization at reduced $|V_M|$ was used to extract $t$, and the tunnel coupling was found to have an exponential dependence on this gate voltage.  Control of $t$ as a function of such a gate voltage is an important component in a number of approaches to quantum dot spin qubits.

We acknowledge useful discussions about quantum dot gate geometries with L. M. K. Vandersypen and C. M. Marcus.  This work was supported in part by ARO and LPS (W911NF-08-1-0482), by NSF (DMR-0805045), by DOD, and by DOE (DE-FG02-03ER46028).  The views and conclusions contained in this document are those of the authors and should not be interpreted as representing the official policies, either expressly or implied, of the U.S. Government.  This research utilized NSF-supported shared facilities at the University of Wisconsin-Madison.

\bibliographystyle{apsrev}
\bibliography{C.B.Simmons_Bib}
\end{document}